\documentclass[12pt]{article}
\usepackage[a4paper,margin = 2cm]{geometry}
\usepackage[T1]{fontenc}
\usepackage{times}
\usepackage{amsmath}
\usepackage{amsfonts}
\usepackage{setspace}
\usepackage{newtxtext,newtxmath}
\usepackage{graphicx}
\usepackage{rotating}
\usepackage{authblk}
\usepackage{color}
\usepackage{subfig}
\usepackage[numbers]{natbib}
\usepackage{tikz}
\usepackage{bm}
\usepackage{bigints}
\usepackage{multirow}
\usepackage{float}
\usepackage{longtable}
\usepackage{lscape}
\usepackage{setspace}

\usetikzlibrary{shapes,decorations,arrows,calc,arrows.meta,fit,arrows,shapes.arrows,
shapes.geometric,shapes.multipart,decorations.pathmorphing,positioning,shapes.swigs}
\tikzset{
    -Latex,auto,node distance =1 cm and 1 cm,semithick,
    state/.style ={ellipse, draw, minimum width = 0.7 cm},
    point/.style = {circle, draw, inner sep=0.04cm,fill,node contents={}},
    bidirected/.style={Latex-Latex,dashed},
    el/.style = {inner sep=2pt, align=left, sloped}
}

\newcommand{\Esp}{\mbox{E}}

\newcommand\independent{\protect\mathpalette{\protect\independenT}{\perp}}
\def\independenT#1#2{\mathrel{\rlap{$#1#2$}\mkern2mu{#1#2}}}

\title{Challenges in systematic reviews and meta-analyses of mediation analyses}

\author[1,2]{Tat-Thang Vo}
\author[2,3]{Stijn Vansteelandt}
\affil[1]{Department of Statistics, The Wharton School, University of Pennsylvania, USA}
\affil[2]{Department of Applied Mathematics, Computer Science and Statistics, Ghent University, Belgium}
\affil[3]{Department of Medical Statistics, London School of Hygiene and Tropical Medicine, UK}

\begin{document}
\maketitle

\abstract{Systematic reviews and meta-analyses of mediation studies are increasingly being implemented in practice. Nonetheless, the methodology for conducting such review and analysis is still in a development phase, with much room for improvement. In this paper, we highlight and discuss challenges that investigators face in mediation systematic reviews and meta-analyses, then propose ways of accommodating these in practice.}

\setlength{\baselineskip}{24pt}

\begin{center} \textbf{Full-text word count}: 3974 \end{center}
\section{Introduction}
Mediation analysis is a common type of statistical analysis in psychology, sociology, epidemiology and medicine. Such analysis aims at assessing the relative magnitude of different pathways and mechanisms by which a treatment may affect an outcome \citep{vanderweele2016mediation, vo2020conduct}. Recently, several systematic reviews and meta-analyses of mediation analyses have been conducted \citep{lubans2008review, gu2015mindfulness, mansell2013and, lee2015does}. These summarize the available evidence regarding the potential mechanisms that underpin the effect of a treatment, thereby refining treatments to improve health outcomes and facilitating the translation of research findings into clinical practice and policy \citep{cashin2019overview}. As for any medical systematic review, a mediation systematic review starts by an extensive search of all studies evaluating the mechanisms of the treatment of interest, followed by an assessment of the similarity of these studies in terms of Population, Intervention, Control and Outcome (PICO). Eligible studies are then assessed in terms of the risk of bias, and those that are sufficiently similar can be meta-analyzed to produce a summary effect estimate \citep{higgins2019cochrane}. A mediation systematic review, however, is often more complicated as it focuses on synthesizing indirect treatment effects. These complications include a high risk of publication bias and selective reporting of mediation results, the lack of a valid tool to evaluate the risk of bias in mediation analyses, heterogeneity across eligible studies in mediator-outcome confounding adjustment and in mediator/outcome measurement. We discuss each of these issues in detail below, and propose potential solutions for overcoming them in practice.
\section{Challenges in systematic reviews of mediation studies}
\subsection{Pros and cons of different searching strategies}
\begin{figure}[H]
\centering
\captionsetup{margin=1.5cm, font={stretch=1.5}}
\begin{tikzpicture}[node distance =1 cm and 1 cm]
    \node[] (x) at (0,0) {$X$};
    \node[] (y) at (2.5, 0)   {$Y$};
    \node[] (m) at (1.15, 1.25)   {$M$};
    \node[] (c) at (2.5, 1.25)   {$L$};

    \path (x) edge (y); \path (x) edge (m);\path (m) edge (y);\path (c) edge (y); \path (c) edge (m);
\end{tikzpicture}

    
    
\caption{The underlying causal graph in each mediation study in the meta-analysis (assuming that the treatment is randomized (1:1) within each eligible study)}
\label{fig1}
\end{figure}
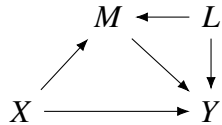

In practice, the aim of a mediation systematic review could be either to summarize all evidence regarding the different pathways that can explain the treatment effect, or to focus particularly on one or certain pathways via some pre-specified mediators of interest. For the sake of simplicity, we will discuss hereafter the simplest (nonetheless most common) setting in which the review is conducted to synthesize the evidence regarding the mediating role of a single continuous variable $M$ in explaining the causal relationship between a randomized treatment $X$ and a continuous outcome $Y$ (figure \ref{fig1}). $M$ and $Y$ obey simple linear models. In such case, one possible strategy is to search for all studies assessing the causal relationship among any two variables (i.e. $X-M$, $M-Y$ and $X-Y$), then quantitatively summarizing these bidirectional findings by a statistical approach such as correlation-based structural equation modeling (see the following section). Such approach does not restrict the eligibility criteria of the systematic review to the studies that considered a formal mediation analysis. This allows one to extensively identify any piece of evidence that can be useful to support the absence or presence of an indirect effect via $M$. This is important. Assume for instance that a large number of studies in which no mediation analysis was performed, found a near null causal effect of the treatment $X$ on the mediator $M$. These studies are at least partially informative as they indirectly suggest the absence of an (important) mediated effect via $M$. In many cases, being aware of such studies can be of critical importance, especially when most mediation analyses identified in the literature suggest (contrarily) a non-null indirect effect via $M$ (e.g. as a result of selective reporting bias). The challenge of this searching strategy, however, is that it may require substantially more time and effort to conduct the review. For the simple setting of single mediator, at least two independent systematic reviews will need to be implemented to retrieve studies investigating the causal effect of $X$ on $M$ and of $M$ and $Y$. These bidirectional studies can also be quite heterogeneous among themselves. The mediator $M$ might be measured by different scale or at different time point in $X-M$ and $M-Y$ studies, hence lumping all these different measurements of $M$ into one variable in the analysis seems problematic. The temporal aspect among the treatment, mediator and outcome in the causal graph is also hard to ensure, as each bidirectional relationship is assessed separately in independent studies.

Alternatively, one could only search for just the individual studies that conduct a formal mediation analysis to assess the indirect effect of interest. This has the advantage that the treatment, mediator and outcome are jointly investigated within each eligible study. Given that these individual mediation studies are properly designed and conducted, certain epidemiological aspects of a mediation analysis (i.e. the temporal order or the treatment, mediator and outcome) can be ensured by the study investigators or further assessed for validity by the reviewer. The practical challenge of this approach, however, is that  mediation analysis is often conducted as a secondary analysis, which makes them more prone to bias. Vo et al (2020) found that 52\% of recent mediation studies on MEDLINE were secondary analyses of previously published RCTs \citep{vo2020conduct}. Similarily, Murillo et al (2021) reported a percentage of about 82\% of studies investigating the mechanisms of cognitive behavioral therapies in chronic pain being conducted secondarily after the publication of the primary trial \citep{murillo2021how}. As these mediation analyses were not pre-planned, it raises concerns over the risk of selectively reporting the mediation results based on statistical significance or in favor of any proposed hypothesis. For instance, inverstigators may intentionally analyze the outcome and mediators on different measurement scales, develop multiple analyses corresponding to multiple causal graphs or path diagrams which assume different causal relationships among the variables, adjust for different sets of mediator-outcome confounders, or implement a series of single mediation analyses but only report those with statistically significant indirect effects in the publication. As a consequence, a mediation systematic review summarizing the available literature may be subject to a high risk of selective reporting bias. Moreover, recent evidence also shows that most mediation analyses are more likely implemented when the primary intent-to-treat (ITT) analysis for the outcome shows statistical significance \citep{murillo2021how}. Without a non-null ITT total effect, mediation analyses are less likely reported as they are anyway not pre-planned. Such practice also biases mediation results toward positive findings. 

In a recent (ongoing) overview of mediation systematic reviews, we find that the second searching strategy is more often adopted in practice. To minimize the risk of selective reporting bias with such a strategy, mediation analyses are ideally pre-planned in the study protocol, including the information on the (measurement of) treatment, the mediators and the outcome of interest, as well as the statistical plan for data analysis. Such protocol should be accessible on public registry platforms as for standard trials and observational studies, to enable subsequent reviewers to assess the risk of negative mediation findings not being reported in the literature.

\subsection{The absence of a valid quality assessment tool for mediation studies}
After selecting eligible studies, the next step in any systematic review and meta-analysis concerns the risk of bias assessment of the individual studies. While several consensus-based bias assessment tools have been proposed to assess the risk of bias in general RCTs (e.g. RoB 2.0 tool \citep{higgins2016revised}) and observational studies (e.g. ROBINS-I tool \citep{sterne2016robins}), to the best of our knowledge, no tool as such has been developed for trials or observational studies with a mediation analysis. Such tool, however, is neccessary as mediation findings are often subject to specific biases. These include temporal order bias which may arise when mediators are not measured prior to the outcome and posterior to the treatment completion, mediator-outcome confounding bias which may arise even in high quality randomized trials, and the use of an inappropriate mediation method (e.g. the traditional product-of-coefficient approach) in the presence of treatment-mediator interactions or when the outcome and/or mediator of interest obey a non-linear model \citep{vanderweele2016mediation, fairchild2017best}. In practice, most mediation literature reviews are either not evaluating mediation-related bias or evaluate them by simple checklists suggested by one or a few experts \citep{lubans2008review, gu2015mindfulness, mansell2013and, lee2015does}. Despite their usefulness, these checklists need to be further improved and extended, for instance, by clearly distinguishing between methodological bias (that impacts the internal validity of the findings) and other reporting issues (e.g. whether the eligible studies cited a theoretical framework to justify the conducted mediation analysis or reported participant flow) \citep{gu2015mindfulness,lubans2008review,mansell2013and}. In future works, these checklists can also be used as a starting point to construct a valid bias assessment tool for mediation analysis. Such a tool should be consensus-based, rigorously developed and validated, and followed by a clear, comprehensive guidance for practical applications. This will be a critically important step towards a better quality of mediation systematic reviews in future practice.  
\section{Challenges in meta-analysis of mediation studies}
In mediation systematic reviews, deriving a summary indirect effect estimate by a meta-analytic approach is possible when the eligible studies are sufficiently similar regarding the population, treatments, mediator and outcome of interest. In this section, we will briefly review the different meta-analysis approaches proposed in the literature to summarize the mediation aggregated data across studies, as well as their advantages and limitations. As no individual participant data (IPD) meta-analysis approach for mediation has been proposed in the literature, we will suggest a simple counterfactual-based approach to meta-analyze the IPD of different mediation analyses. This approach extends current works on causally interpretable meta-analysis to a mediation setting. In contrast to other proposed approaches, it has the advantage of being explicit about the target population for which the summary estimate describe the indirect treatment effect.   
\subsection{Mediation meta-analysis approaches using aggregated data: a brief overview}
We concentrate on the simplest setting of single mediation analysis (figure \ref{fig1}) with a randomized treatment $X$ and no mediator-outcome confounder ($L$ is an empty set). The complexity due to mediator-outcome confounders will be discussed in the next section. In practice, many different meta-analysis approaches have been proposed to summarize the aggregated mediation data \citep{cheung2016random}. The choice of approach will depend upon which of the two searching strategies is used to identify eligible studies.

The so-called parameter-based meta-analytic structural equation modeling (MASEM) approach is used when the review only includes studies which perform a formal mediation analysis. In this approach, the estimate of the indirect effect via $M$ is first extracted from the publication of each individual mediation study. The resulting indirect effect estimates are then meta-analyzed by a standard random-effect meta-analysis model \citep{cheung2016random}, which can be expressed as:
\[ \hat\theta_i = \theta + u_i + e_{i}\]
where $\hat\theta_i$ denotes the indirect effect estimate from study $i$ and $\theta_i$ denotes the summary indirect effect. The between- and within-study heterogeneity across studies are reflected through the random effect $u_i \sim \mathcal{N}(0,\tau^2)$ and the random error $e_i\sim \mathcal{N}(0,\sigma_i^2)$ with $\sigma_i^2$ known, respectively. A Bayesian extension of this approach has also been proposed in the literature \citep{van2020comparison}. In practice, researchers sometimes meta-analyze additionally the $X-M$ and $M-Y$ (conditioning on $X$) relationships if the individual mediation studies also provide the corresponding effect estimates. Results of these secondary meta-analyses, however, might be subject to bias as they may not take into account all relevant studies evaluating the bidirectional $X-M$ and $M-Y$ relationships in the literature, as some may not have performed a mediation analysis. Moreover, the product of these path-specific summary estimates (denoted $\overline{a}$ and $\overline{b}$) does not necessarily equal the weighted average $\overline{ab}$ of multiple products of coefficients produced in the individual studies. 

The so-called correlation-based MASEM approach is used when all bidirectional studies from the literature assessing the impact of (i) $X$ on $Y$ or $M$ and of (ii) $M$ on $Y$ are identified and included in the analysis. In this approach, a multivariate random-effect model is first fitted to summarize the correlation matrices of the different variables involved in the mediation process. In the considered setting (figure 1 with $L$ be an empty set), this model can be expressed as:
\[\bm r_i = \bm\rho + \bm u_i + \bm e_{i}\]
where $\bm r_i = \begin{pmatrix} \hat r_{XY} & \hat r_{XM} & \hat r_{MY}\end{pmatrix}^T$ is the $3\times1$ sample correlation vector obtained from study $i$. Here, $\hat r_{XY}$, $\hat r_{XM}$ and $\hat r_{MY}$ denotes the pairwise (marginal) correlation coefficient estimate of $X$ and $Y$, of $X$ and $M$ and of $M$ and $Y$, respectively. $\bm \rho= \begin{pmatrix} r_{XY} & r_{XM} & r_{MY}\end{pmatrix}^T$ denotes the $3\times 1$ average correlation vector. The vector of random effects $\bm u_i\sim \mathcal{N}(\bm 0,\mathrm{\textbf{T}})$ captures the difference across studies in the population correlation coefficients and the residual vector $\bm e_i\sim \mathcal{N}(\bm 0,\bm \Sigma_i)$ with $\Sigma_i$ known captures the sampling error. This approach can handle missing correlation coefficients in studies that for instance only evaluated the $X-M$ association by assuming that mising correlation coefficient estimates in a study are either missing completely at random (MCAR) or missing at random (MAR) \citep{cheung2016random}. Once the correlation model is fitted, the estimated average correlation matrix $\bm{\hat\rho}$ and the corresponding (estimated) covariance matrix $\bm V$ are used to fit the proposed structural mediation model (figure 1) \citep{cheung2016random}. To achieve this, consider $\bm\rho$ as a function of the parameters $\bm\theta$ indexing the structural mediation model 1, i.e. $\bm\rho=\bm\rho(\bm \theta)$. Cheung (2014) proposed using the weighted least squares (WLS) estimation method to fit this model \citep{cheung2014fixed}. The discrepancy function that describes how closely the structural model 1a conforms to the observed data can be expressed as:
\[F(\bm\theta) = (\bm{\hat\rho} - \bm\rho(\bm\theta))^TV^{-1} (\bm{\hat\rho} - \bm\rho(\bm\theta))\]
A likelihood ratio test and various goodness-of-fit indices can be used to evaluate the appropriateness of the proposed mediation model.

Apart from the MASEMs, a so-called marginal likelihood (ML) approach has also been proposed to meta-analyze studies using product-of-coefficient approach to investigate mediation \citep{huang2016statistical}. Here and below, denote $S$ the trial from which a patient originates. $S$ takes value from $1$ to $K$, where $K$ is the number of studies in the meta-analysis. Assume that $\Esp(M|X,S=i) = \alpha_{0i} + \alpha_iX$ and $\Esp(Y|X,M,S=i) = \beta_{0i} + \beta_{1i}X + \beta_iM$. A random-effect model is then imposed on the estimates $a_i$ and $b_i$ of $\alpha_i$ and $b_i$, respectively. More precisely:
\begin{align*}
a_i \sim \mathcal{N}(\alpha_i,\psi_i^2);~~~~~~~~ b_i \sim \mathcal{N}(\beta_i,\phi_i^2);~~~~~~~~\begin{pmatrix} \alpha_i \\ \beta_i \end{pmatrix}\sim \mathcal{N}\bigg[\begin{pmatrix} \mu_\alpha \\ \mu_\beta \end{pmatrix}, 
\begin{pmatrix} \sigma_{11} & \sigma_{12} \\ \sigma_{12} & \sigma_{22} \end{pmatrix}
\bigg]
\end{align*} 
with $\sigma_{ij}$ $(i,j=1,2)$ known. Such a model can be estimated by using (restricted) maximum likelihood, which then allows one to estimate the target parameter $\delta = \mu_\alpha\mu_\beta$ by $\hat\delta = \hat\mu_\alpha\hat\mu_\beta$, where $ \hat\mu_\alpha$ and $\hat\mu_\beta$ are the likelihood-based estimates of $\mu_\alpha$ and $\mu_\beta$. The standard error SE$(\hat\delta)$ of $\hat\delta$ is estimated using Sobel's formula (1982), i.e. $ \hat{\mathrm{SE}}(\hat\delta) = \sqrt{ \hat\mu_\alpha^2s^2_{\hat\mu_\beta} + \hat\mu_\beta^2s^2_{\hat\mu_\alpha}}$ where $s^2_{\hat\mu_\beta}$ and $s^2_{\hat\mu_\alpha}$ are the estimates of the variance of $\hat\mu_\beta$ and $\hat\mu_\alpha$, respectively. As $\hat\delta$ is typically not normally distributed, bootstrap methods are often used to construct the 95\% confidence interval of $\delta$ \citep{van2020comparison, huang2016statistical}. A Bayesian extension of this approach has also been proposed in the literature \citep{van2020comparison}.
\subsection{Methodological concerns in mediation meta-analysis with aggregated data}    
\textbf{Complexity due to mediator-outcome confounders} --  In the correlation-based MASEM approach, confounding adjustment is not readily allowed for. Practical implementations often implicitly assume that there is no mediator-outcome confounding other than by treatment itself, which is quite unlikely in practice \citep{cheung2016random}. In the parameter-based MASEM and the marginal likelihood approach, confounding adjustment is essentially considered in each eligible mediation study. The challenge, however, is that different studies may adjust for a different set of mediator-outcome confounders. In practice, about 40\% of mediation studies do not adjust for mediator-outcome confounders. The other 60\% often adjust for a quite small set of confounders, which in most cases only includes the baseline values of the mediator and outcome. Few studies adjust for a large, extensive set of confounders  \citep{vo2020conduct}. When the outcome or mediator of interest is binary or time-to-event, such heterogeneity in confounding adjustment can be even more problematic due to non-collapsibility of the odds ratio or hazard ratio \citep{vanderweele2016mediation}. Unfortunately, relying on the aggregated-data from the eligible studies as in the aforementioned approaches can hardly address this challenge. The issue is analogous to that in meta-analysis of standard observational studies, where diffferent studies may consider different sets of exposure-outcome confounders \citep{metelli2020challenges, liu2017can}. To overcome the challenge, individual participant data (IPD) meta-analysis approaches are needed \citep{6debray2015get}. Accessibility of IPD allows one to standardize the covariate adjustment across studies, and to apply more appropriate or advanced methods when necessary. For instance, when the data on some important mediator-outcome confounders are collected in some but not all studies, imputation methods may be needed to impute these systematically missing covariate data, based on what is observed in other studies with available information on these covariates \citep{quartagno2016}.

\textbf{Complexity due to non-linearities} -- When the mediator and outcome of interest are both continuous and obey simple linear (main effect) models, the parameter-based MASEM and marginal likelihood approaches can be used to derive a summary (natural) indirect effect estimate across multiple populations, given that the mediator-outcome confounders are properly taken into account. When the mediator or outcome obey nonlinear models (e.g. binary mediator/outcome and/or treatment-mediator interaction), then valid analogs to the above approaches are more challenging to develop and have not been worked out to the best of our knowledge. However, if the main interest in each eligible study is to test for the evidence of an indirect effect via $M$, then the product-of-coefficient method remains valid \citep{vanderweele2016mediation}. Even the difference-of-coefficient method could be used for a conservative test and non-collapsibility is not a concern (see for instance, the discussion by Jiang and VanderWeele \citep{jiang2015difference}). Future research should therefore primarily focus on answering how a mediation meta-analysis could be developed (and whether the above methods apply) with the sole aim to do mediation testing in the presence of non-linearities. 

When the aim of the meta-analysis is to quantify the magnitude of the (summary) indirect effect then in the absence of individual participant data, more complications arise as the product-of-coefficient approach may no longer provide a proper estimate for the natural indirect effect within each mediation study. Simply meta-analyzing multiple product-of-coefficient estimates as in the parameter-based MASEM approach will make the final summary estimate hard to interpret. Similar concerns apply to the correlation-based MASEM an the marginal likelihood approach.

\textbf{Complexity due to measurement scale heterogeneity} -- One other practical problem that one may have to deal with in a mediation meta-analysis is the use of different measurement scales of the mediator and outcome across studies, which is particularly common in psychological research. Accounting for this scale difference is however complicated, because most studies often use only one specific scale to measure each variable. In the correlation-based MASEM approach, the correlation data are summarized instead of the covariance data to avoid the impact of such measurement scale heterogeneity. In the parameter-based MASEM approach, the regression coefficients or the product of regression coefficients in each individual study are often standardized by dividing the original coefficients and products by the standard deviation of the corresponding variable, so as to account for the scale heterogeneity across studies \citep{cheung2016random}. However, the standard deviation of a variable is extremely sensitive to arbitrary features of a study's design. This makes standardized coefficients (and standardized products of coefficients) vary capriciously with design factors such as admission criteria of the study, while unstandardized coefficients are not influenced by such factors (see a toy example in appendix 1). Several experts hence discourage the use of standardized coefficients in epidemiological practice, especially when the interest is in comparing the observed effect of one variable on another across different settings \citep{greenland1991standardized,greenland1986fallacy}. Alternatively, individual studies should be encouraged to measure subjective mediators and outcomes by multiple scales, then reporting the concordance between the two scales (e.g. by calculating correlation coefficient) as part of the analysis results. When there is partial overlap between mediation studies regarding these measurement scales, for instance, the mediator of interest is measured by scale $A$ and $B$ in one study, but by scale $B$ and $C$ in another study, future research could aim to investigate whether a factor analysis could be considered to take into account this scale heterogeneity when meta-analyzing the mediation findings. To make this strategy become more feasible in practice, it would be appealing to establish a list of recommended measurement scales for the variables that are commonly of interest in a given research field. Investigators might then choose among these recommended scales the ones most appropriate for their research setting.

As a final remark, it is worth noting that the performance of the aforementioned mediation meta-analysis approaches has recently been compared by simulated data \citep{van2020comparison}. This simulation study, however, considers the simplest setting of simple linear mediator and outcome models, without mediator-outcome confounding and/or treatment-mediator interaction in the outcome generating mechanism. Future research should hence focus on evaluating the performance and validity of these approaches in more complex but realistic settings, where the above difficulties present.  
\subsection{Mediation meta-analysis with individual participant data}
As in standard meta-analysis, the mediation meta-analysis approaches discussed above make use of random-effect models to express heterogeneity across studies. These models interpret a weighted average of study results as an estimate of a mean parameter across a hypothetical population of studies. The relevance of this methodology to patient care is not evident, as clinicians need to assess treatments and their mechanisms for populations of patients, not for populations of studies \citep{manski2020toward}. Recently, several new approaches have been proposed to overcome this challenge. These approaches aim to standardize results of different studies over the case-mix of a well-defined target population, prior to applying conventional meta-analyis techniques to summarize thes findings \citep{sobel2017causal, vo2019rethinking, dahabreh2020toward}. In this section, we describe the extension of these approaches to meta-analysis of mediation studies. 

We aim to estimate the indirect effect via the mediator candidate $M$ if a given mediation analysis were conducted in an external population different from the actual study population. When all indirect effect estimates are standardized over the same population, the subsequent meta-analysis will derive a summary indirect effect estimate that quantifies the role of the mediator $M$ in explaining the treatment effect in the target population of interest.

\begin{figure}[H]
\centering
\captionsetup{margin=1.5cm, font={stretch=1.2}}
\begin{tikzpicture}[node distance =1 cm and 1 cm]
    \node[state,draw = none] (x) at (0,0) {$X$};

    \node[state,circle, draw = white] (sp) [right =of x] {$M$};
    \node[state,circle, draw = white] (y) [right =of sp] {$Y$};
 
    \node[state,circle, draw = white] (s) [above =of sp] {$S$};
    \node[state,circle, draw = white] (sp1) [right =of s] {$ $};
    \node[state,circle, draw = white] (l) [above =of sp1] {$L$};
    \node[state,circle, draw = black] (ul) [right =of sp1] {$U_l$};
  
    \path (x) edge (sp); \path(sp) edge (y); \path (s) edge (x);\path (s) edge (y);\path (l) edge (s);\path (l) edge (y); \path (s) edge (sp);
    \path (ul) edge (l); \path (ul) edge (y);
    \path (l) edge (sp);
    \draw[->](x) to[out=-35,in=-145] (y);
\end{tikzpicture}
\caption {A causal diagram illustrating the setting of interest}
\end{figure}
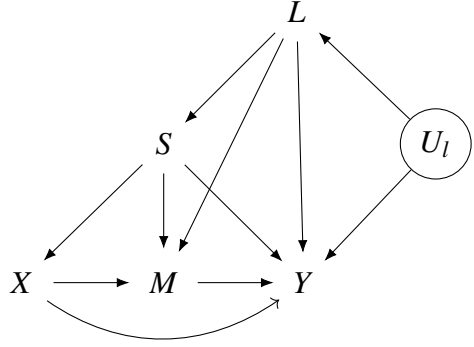

To achieve this, assume that the causal diagram depicted in figure 2 describes the causal relationships between the different variables. As the treatment versions can vary across different studies, we denote $x_k~ (x=0,1;~k=1\ldots K)$ as the treatment version used in study $S=k$. Besides, denote $Y(x_k,M(x_k^*))$ the nested counterfactual outcome that could be observed if a patient were treated by the treatment version $x_k$, fixing the mediator $M$ at the value potentially observed under the treatment version $x_k^*$. The natural indirect effect estimated in study $k$ if this study were instead conducted in population $S=j$ can be defined as:
\begin{align}
\theta(j,k) =  \Esp(Y(x_k,M(x_k^*))|S=j) - \Esp(Y(x_k,M(x_k))|S=j).
\end{align}
The estimation of this indirect effect hence relies on the identifiability of the expectation $\Esp(Y(x_k,M(x_k^*))|S=j)$ ( $x,x^*=0,1$). The causal assumptions under which such expectation can be estimated from data are formally stated in appendix 2. Apart from the standard assumptions often made in mediation analysis such as the "cross-world" assumption and the absence of unmeasured mediator-outcome confounders within each study, an important requirement is that the set of baseline covariate $L$ includes all outcome and mediator predictors that are differentially distributed across studies. When the data on some components of $L$ are systematically missing in some studies, imputation methods such as the one proposed by \citep{quartagno2016} can be used to impute the values of these covariates, based on what is observed in other studies with these covariates being measured.

Consider now the simplest setting in which the mediator and outcome are both continuous and obey linear (main-effect) models, that is:
\[\Esp(M|X,L,S=k) = \alpha_{0k} + \alpha_{1k}X + \alpha_{2k}L\]
\[\Esp(Y|X,M,L,S=k) = \beta_{0k} + \beta_{1k}X + \beta_{2k}M + \beta_{3k}L\]
In such case, we show in appendix 2 that $\theta(j,k) = \theta(k,k) = \alpha_{1k}\beta_{2k}(x^*-x)$. This implies that $\theta(j,k)$ can be validly estimated by the standard product-of-coefficient approach using data from study $k$, and no adjustment is needed to account for the difference in the covariate distribution between the two population $j$ and $k$. In more general settings where non-linearities present, the above equality will no longer hold and novel estimating strategies are required. A simple approach could be to postulate and fit a natural effect model conditional on $L$ within study $k$, then use that model to predict the countefactual outcome $Y_i(x_k,M(x_k^*))$ for each patient $i$ in study $j$ \citep{steen2017medflex}. The average of these predicted outcome values will provide a valid estimate for $ \Esp(Y(x_k,M(x_k^*))|S=j)$ (appendix 2). 

In what follows, the case-mix standardized estimates $\hat\theta(j,k)$ can be summarized by the standard random-effect meta-analysis model. More precisely, a population-$j$-specific meta-analysis summarizing different $\hat\theta(j,k)$'s with the same index $j$ can be fitted, for then $\hat\theta(j,k) = \theta + \alpha_k + \epsilon_k$, with $\alpha_k \sim \mathcal{N}(0,\tau_j^2)$ denoting the random effect and $\epsilon_k \sim \mathcal{N}(0,\sigma^2_k)$ (with $\sigma^2_k$ known) denoting the random error. The fixed-effect component $\theta_j$ in this model describes the summary indirect effect of $X$ on $Y$ via the mediator $M$ in the target population $j$, and the variance component $\tau_j^2$ describes the heterogeneity between the individual indirect effect effects across studies even after being standardized over the same covariate distribution of the trial population$j$. Alternative random-effect models for summarizing $\hat\theta(j,k)$ can also be extended from previous works \citep{vo2021assessing}. 

The above proposal is recommended when all mediation analyses investigating the mediating role of $M$ in explaining the $X-Y$ association are retrieved from the literature. As discussed earlier, in certain cases, studies that only investigate the impact of $X$ on $M$ could also be (partially) informative when assessing the indirect effect via $M$. In appendix 3, we thus propose an approach to integrate the $X-M$ studies into the meta-analysis of (formal) mediation studies, taking into account the difference in the target population between these studies. This approach, however, assumes that controlling for the baseline covariates in $X-M$ studies would be sufficient to adjust for the mediator-outcome confounders if these studies implemented a formal mediation analysis as in the mediation studies.  Imputation methods will otherwise be needed to impute the missing covariate data.   
\section{Conclusion}
In this paper, we highlight and discuss many potential challenges that one may have to deal with when conducting mediation systematic reviews and meta-analyses. These challenges should be further addressed in future works to satisfy the increasing need of summarizing mediation evidence in applied research practice. In the second part of this paper, we also propose a simple approach to standardize results of different mediation studies over the case-mix of a target population before meta-analyzing them, so as for the final summary indirect effect estimate to have a causal interpretation. This approach will be better applicable when current mediation practice is further improved, especially regarding mediator-outcome confounding adjustment and mediator/outcome measurement. 

\bibliographystyle{wileyj}
\bibliography{library}
\section*{Appendix 1}
We here provide a toy example to illustrate the limitation of standardized product-of-coefficients estimates in quantifying the indirect effect. This example is motivated by the discussion of Greenland et al (1991) \citep{greenland1991standardized}. Consider two studies ($S=1,2$) assessing the mediating role of a continuous mediator $M$ in explaining the relationship between blood pressure ($X$) and quality of life ($Y$). Assume further that $\Esp(M|X,S=i) = \alpha_{0i} + \alpha X$ and $\Esp(Y|X,M,S=i) = \beta_{0i} + \beta_{1i}X + \beta M$, where $i=1,2$. The indirect effect (IE) will thus be the same across two studies, i.e. $\alpha\beta$. When the sample sizes of these studies are sufficiently large, the two unstandardized IE estimates (i.e. $a_ib_i$ where $a_i$ and $b_i$ are estimates of $\alpha$ and $\beta$ using data from study $i$) are expected to be sufficiently close. 

The two standardized IE estimates, however, will equal $a_ib_i \hat{\sigma}(X|S=i)/\hat{\sigma}(Y|S=i)$, where $\hat{\sigma}(X|S=i)$ and $\hat{\sigma}(Y|S=i)$ are observed standard deviation of $X$ and $Y$ in study $i$ ($i=1,2$). Now assume that patients in study $S=1$ are mostly elderly (e.g. 60 -- 80 years of age), while patients in study $S=2$ are more heterogeneous in age (e.g. 40 -- 80 years). Due to this design-related factor, we may observe a higher variation in the blood pressure among patients in study $S=2$ than in study $S=1$. As a result, the standardized IE estimate in study $S=2$ will potentially be larger than in study $S=1$, although we there is no stronger mediation evidence in study $S=2$. 
 
\section*{Appendix 2}
The assumptions under which the parameter $\theta(j,k)$ can be estimated from the data are provided below:
\begin{enumerate}
\itemsep0em 
\item [(i)] \textbf{Consistency}, i.e. $\Pr(M(x_k)=M|X=x,S=k)=1$ and $\Pr(Y(x_k,m)=Y|X=x,M=m,S=k)=1$
\item [(ii)] \textbf{Positivity}, i.e. $\Pr(0 < \Pr(S=j|L) < 1)=1 ~\forall j$ and $\Pr(0 < \Pr(M|L,X=x,S=j) < 1)=1 ~\forall x,j$, which ensures that studies are sufficiently similar in terms of baseline covariate and mediator distribution to avoid unreasonable extrapolations
\item [(iii)] \textbf{Within-trial ignorability}, i.e. $M(x_k)\independent X|L,S$ and $Y(x_k,m)\independent X|L,S$, which is often satisfied in the context of randomized trials
\item [(iv)] \textbf{Between-trial ignorability}, i.e. $Y(x_k,m)\independent S| X=x,L$ and $M(x_k) \independent S|L,X=x$, which requires that the set of baseline covariate $L$ include all outcome and mediator predictors that are differentially distributed across studies
\item [(v)] \textbf{No unmeasured mediator-outcome confounders within each study}, i.e. $Y(x_k,m)\independent M| X=x,L,S=k$
\item [(vi)] \textbf{Within-study cross-world assumption}, i.e. $Y(x_k,m) \independent M(x_k^*)|L,S$. This assumption generalizes the standard cross-world assumption often made in mediation analysis, taking into account the presence of multiple treatment versions across studies. Such assumption is satisfied under the non-parametric structural equation model with independent errors associated with the causal diagram depicted in figure 2.  
\end{enumerate}
Under the above assumptions, one then has:
\begin{align*}
 \Esp(&Y(x_k,M(x_k^*))|S=j)\\
&=\Esp\{  \Esp(Y(x_k,M(x_k^*))|L,S=j)|S=j\}\\
&=\Esp\left\{ \int \Esp(Y(x_k,m)|L,S=j)\cdot f_{M(x_k^*)}(m|L,S=j)dm\bigg\vert S=j \right\}\\
&=\Esp\left\{ \int \Esp(Y|X=x,M=m,L,S=k)\cdot f_M(m|X=x^*,L,S=k)dm\bigg\vert S=j \right\}\\
&=\Esp\{ \Esp(Y(x_k,M(x^*_k))|L,S=k)|S=j\}\\
&= \Esp \{ I(S=j)\cdot\Esp(Y(x_k,M(x^*_k))|L,S=k) \}\cdot \Pr(S=j)^{-1}
\end{align*}
This suggests a simple approach to estimate $\Esp(Y(x_k,M(x_k^*))|S=j)$, which is to postulate a natural effect model of the form:
\begin{align*}\Esp(Y(x_k,M(x^*_k))|L,S=k) = \beta_{0k} + \beta_{1k}x + \beta_{2k}x^* + \beta_{3k}L\end{align*}
then estimating $ \Esp(Y(x_k,M(x_k^*))|S=j)$ as:
\begin{align*}
\hat \Esp(&Y(x_k,M(x_k^*))|S=j) = \frac{1}{N_j} \cdot \sum_i I(S_i=j)\cdot (\hat \beta_{0k} + \hat\beta_{1k}x + \hat\beta_{2k}x^* + \hat\beta_{3k}L_i ) 
\end{align*}
where $N_j$ is the total number of patients in study $j$.

We now develop more insight for the simplest setting in which  the mediator and outcome are continuous and obey linear (main effect) model, that is:  
\[\Esp(M|X,L,S=k) = \alpha_{0k} + \alpha_{1k}X + \alpha_{2k}L\]
\[\Esp(Y|X,M,L,S=k) = \beta_{0k} + \beta_{1k}X + \beta_{2k}M + \beta_{3k}L\]
In this case, one can further show that $\theta(j,k) = \theta(k,k) = \alpha_{1k}\beta_{2k}(x^*-x)$, which implies that $\theta(j,k)$ can be validly estimated by the standard product-of-coefficient approach using data from study $k$, and no adjustment is needed to account for the difference in the covariate distribution between the two population $j$ and $k$. Indeed,
\begin{align*}
 \Esp(&Y(x_k,M(x_k^*))|S=j)\\
&=\Esp\left\{ \int \Esp(Y|X=x,M=m,L,S=k)\cdot f_M(m|X=x^*,L,S=k)dm\bigg\vert S=j \right\}\\
&=\Esp\left\{ \int (\beta_{0k} + \beta_{1k}x + \beta_{2k}m + \beta_{3k}L)\cdot f_M(m|X=x^*,L,S=k)dm\bigg\vert S=j \right\}\\
&=\beta_{0k} + \beta_{1k}x + \beta_{2k} \int m f_M(m|X=x^*,L,S=k)dm+ \beta_{3k}\Esp\bigg\{ \int L f_M(m|X=x^*,L,S=k)dm\bigg|S=j\bigg\}\\
&=\beta_{0k} + \beta_{1k}x+ \beta_{2k}\Esp[E(M|X=x^*,L,S=k)|S=j]+ \beta_{3k}\int\int lf_M(m|X=x^*,l,S=k)f_L(l|S=j)dmdl\\
&=\beta_{0k} + \beta_{1k}x+ \beta_{2k}(\alpha_{0k} + \alpha_{1k}x^* + \alpha_{2k}\Esp(L|S=j)) +  \beta_{3k} \int l\left(\int f_M(m|X=x^*,l,S=k)dm\right)f_L(l|S=j)dl\\
&=\beta_{0k} + \beta_{1k}x+ \beta_{2k}(\alpha_{0k} + \alpha_{1k}x^* + \alpha_{2k}\Esp(L|S=j)) + \beta_{3k}\Esp(l|S=j)
\end{align*}
As a result, $\theta(j,k) =  \Esp(Y(x_k,M(x_k^*))|S=j) - \Esp(Y(x_k,M(x_k))|S=j) = \alpha_{1k}\beta_{2k}(x^*-x)$.
 
\section*{Appendix 3}
We here propose an approach to integrate results of studies investigating the $X-M$ association in the meta-analysis of mediation studies. Assume that in a randomized trial $S=k$, a mediation analysis is conducted to investigate the mediating role of $M$ in explaining the $X-Y$ relationship (e.g. figure 1). In another randomized trial $S=j$, the causal relationship of $X$ on $M$ is investigated. There is no difference across studies in terms of treatment version. 

One possible question is that "if trial $S=j$ also conducted a mediation analysis with the mediator $M$ and the outcome $Y$ as in trial $S=k$, how could the result of such mediation analysis look like?". Our target is then to estimate the natural indirect effect in study $S=j$:
\begin{align}\label{eq2}
\eta(j) = \Esp(Y(x,M(x^*))|S=j) - \Esp(Y(x,M(x))|S=j)
\end{align}
Alternatively, one could also address the question that "if trial $S=j$ was actually conducted in the target population of trial $k$ and also considered a mediation analysis with the mediator $M$ and the outcome $Y$ as in trial $S=k$, how could the result of such mediation analysis look like?". The focus is thus to estimate the natural indirect effect in study $S=k$ by partially using the external data from trial $j$:
\begin{align}\label{eq3}
\eta(k) = \Esp(Y(x,M(x^*))|S=k) - \Esp(Y(x,M(x))|S=k)
\end{align}
To identify (\ref{eq2}) and (\ref{eq3}), the following set of assumptions is needed:
\begin{enumerate}
\itemsep0em 
\item [(i)] \textbf{Consistency}, i.e. $\Pr(M(x)=M|X=x)=1$ and $\Pr(Y(x,m)=Y|X=x,M=m)=1$
\item [(ii)] \textbf{Positivity}, i.e. $\Pr(0 < \Pr(S=j|L) < 1)=1 ~\forall j$ and $\Pr(0 < \Pr(M|L,X=x,S=j) < 1)=1 ~\forall x,j$
\item [(iii)] \textbf{Within-trial ignorability}, i.e. $M(x)\independent X|L,S$ and $Y(x,m)\independent X|L,S$
\item [(iv)] \textbf{Between-trial ignorability}, i.e. $Y(x,m)\independent S| X=x,L$ and $M(x) \independent S|L,X=x$
\item [(v)] \textbf{No unmeasured mediator-outcome confounders within each study}, i.e. $Y(x,m)\independent M| X=x,L,S=k$
\item [(vi)] \textbf{Within-study cross-world assumption}, i.e. $Y(x,m) \independent M(x^*)|L,S$.   
\end{enumerate}
The interpretation of these assumptions is similar to that of the assumptions proposed in appendix 1. Under these assumptions, one can easily show that:
\begin{align}\label{eq4}
 \Esp(&Y(x,M(x^*))|S=j)\nonumber\\
&=\Esp\{  \Esp(Y(x,M(x^*))|L,S=j)|S=j\}\nonumber\\
&=\Esp\left\{ \int \Esp(Y(x,m)|L,S=j)\cdot f_{M(x^*)}(m|L,S=j)dm\bigg\vert S=j \right\}\nonumber\\
&=\Esp\left\{ \int \Esp(Y(x,m)|X=x,L,S=j)\cdot f_M(m|X=x^*,L,S=j)dm\bigg\vert S=j \right\}\nonumber\\
&=\Esp\left\{ \int \Esp(Y(x,m)|X=x,L,S=k)\cdot f_M(m|X=x^*,L,S=j)dm\bigg\vert S=j \right\}\nonumber\\
&=\Esp\left\{ \int \Esp(Y|X=x,L,M=m,S=k)\cdot f_M(m|X=x^*,L,S=j)dm\bigg\vert S=j \right\}
\end{align}
and similarly:
\begin{align}\label{eq5}
 \Esp(Y(x,M(x^*))|S=k)
=\Esp\left\{ \int \Esp(Y|X=x,L,M=m,S=k)\cdot f_M(m|X=x^*,L,S=j)dm\bigg\vert S=k \right\}
\end{align}
Intuitively, the information about the $M-Y$ association (given $X$) from trial $k$ can be used together with the information about the $X-M$ association in trial $j$ to estimate $ \Esp(Y(x,M(x^*))|S=j)$ and $\Esp(Y(x,M(x^*))|S=k)$. 

In what follows, one can use a fixed-effect meta-analysis model to summarize the different estimates of $\eta(j)$ obtained by using data from trials $S=k_1, k_2, \ldots$ with a formal mediation analysis conducted. The summary estimate from such meta-analysis will reflect the indirect effect of $X$ on $Y$ via $M$ in the target population of trial $j$. Alternatively, one can also meta-analyze the different estimates of $\eta(k)$ obtained by using data from trials $S=j_1, j_2, \ldots$ assessing the $X-M$ association. The summary estimate from such meta-analysis will reflect the indirect effect of $X$ on $Y$ via $M$ in the target population of trial $k$. 

The above proposal will be helpful, for instance, when trial $j$ shows no statistical significance of the $X-M$ relationship (which in turns suggests that there might be no indirect effect via $M$). By using the proposed approaches, the findings from such study can be formally integrated in the meta-analysis.

\textbf{Treatment version heterogeneity} -- Assume now that there is heterogeneity between study $j$ and $k$ regarding the version of treatments being used. Denote $x_k$ ($x=0,1$) as the treatment version used in study $S=k$ and $Y(x_k,M(x^*_j))$ as the value of the outcome that would have been observed had the treatment $X$ been set to the version $x_k$ in trial $k$, fixing the mediator $M$ at the value potentially observed under the treatment version $x^*_j$. The focus is then on estimating the following natural indirect effects:
\begin{align*}
\gamma(j,k) = \Esp(Y(x_k,M(x^*_j))|S=k) - \Esp(Y(x_k,M(x_k))|S=k)\\
\delta(j,k) = \Esp(Y(x_k,M(x^*_j))|S=j) - \Esp(Y(x_k,M(x_k))|S=j)
\end{align*}
To estimate the above estimands from the observed data, we make use of the assumptions (i) to (v) proposed to estimate $\theta(j,k)$ in appendix 1 and the (generalized) cross-world assumption which requires that $Y(x_k,M(x^*_j)) \independent M(x^*_j)|L,S$. Under these assumptions, $\Esp(Y(x_k,M(x^*_j))|S=j) $ and $\Esp(Y(x_k,M(x^*_j))|S=k)$ can be expressed as the right-hand side of equations (\ref{eq4}) and (\ref{eq5}), respectively. This hence suggest simple plug-in approaches to estimate $\gamma(j,k)$ and $\delta(j,k)$.

In what follows, one can use a  random-effect meta-analysis model to summarize the different population-$k$-specific estimates $\hat\gamma(j,k)$ with $j=j_1, j_2, \ldots$ denoting studies having data on the $X-M$ association. The summary estimate from such meta-analysis will reflect the indirect effect of $X$ on $Y$ via $M$ in the target population of trial $j$, acknowledging that there is treatment version heterogeneity across studies. Alternatively, one can also meta-analyze the different population-$j$-specific estimates $\hat\delta(j,k)$ with $S=k_1, k_2, \ldots$ denoting studies having data on the $M-Y$ association conditioning on $X$ (i.e. studies with a formal mediation analysis conducted). The summary estimate from such meta-analysis will reflect the indirect effect of $X$ on $Y$ via $M$ in the target population of trial $j$, also acknowledging that there is treatment version heterogeneity across studies. 
\end{document}